# A TOUR OF THE STUDENT'S E-LEARNING PUDDLE

Vidhu Mitha

Department of Information Technology, SRM University, Chennai, India

vidhu.mitha@gmail.com

*ABSTRACT*

*E-learning has revolutionized our realm in more than just a listable number of ways. But it took a paradigm shift when it entered the threshold of the varsity system. With the prevailing spoon-feeding era, are the students really industry ready? We answer that by confirming a fact: web-based learning has become the oxygen of freshers in the IT Industry instead of the traditional learning done through graduation. Furthermore, are university enforced e-learning assessment systems a true representation of a student's proficiency? This paper is a peep into what web-based e-learning systems are to a student of today's world, by giving an overview of university-level e-learning in India deploying an example from SRM University's organizational framework. It assesses a key e-learning trend, the implementation of which bridges the gap between universities and the industry. It is proposed to provide constructive feedback to the e-learning community and shine some light on areas of scope for future developments.*

*KEYWORDS*

*web-based learning, e-learning, university*

## 1. INTRODUCTION

We see trends move toward a NO-PAPER system in education today. But what is the reality?

Ranging from high schools' e-labs, to finding it's traces in fully developed experts' CVs as nanodegrees, e-learning has found its way in, everywhere including in India with an estimated $3 billion market. With the rapidly changing definition of e-learning today, it has swept countless fields and has a universal set of users starting from middle schoolers who've started submitting their assignments online, to corporate biggies that cannot function without it.

Gone are the days that a student accredited an institution's tutoring for his/her learning. Each student now has a pet go-to e-learning facility to thank for saving their tests and scores! This what tempts us to ask if this was used as a last-minute trick and if it helped? If yes, we will discuss if it is really a true representation of a student's skill as compared to a hardworking offline-study student.

But on the other hand, is this system being utilized to its maxed potential?

In universities, fully equipped hi-tech labs are available for training graduates. During the college years, we are taught the practice of multiple technologies, coding languages and its practical implementations. But the fact remains that over 90% of it, is learned online. Throughout this learning process, students are being asked to line up to a system bound by the university, to complete the courses they have defined for the them. This leads to the said underutilization of e-learning. While sticking to pre-defined university courseware, we lose track of the fast-paced industry's growing needs. This would lead us to the discussion on "bridging the gap".

## 2. THROUGH THE STUDENT LENS

On an average, a graduate level student uses books less than once a week. But online content is accessed at least 4-5 days a week and multiple times during a study day. The virtual notebook has taken over the blackboard. A few scenarios can help explain how:

When a student does not know a common definition or is unaware of the meaning of certain phrases or terminology used somewhere, he no longer looks for an encyclopedia, but looks up his phone's google dictionary instead. It is a lot faster than manual searching since it's just a few clicks away.

Not all the students understand at their very first attempt at learning, but most of them wouldn't want to fail in the presence of a class full of students. Web-based learning personalizes the environment, hence no fear to attempt again till they learn it. This enables the student to explore, test their ideas well and most importantly, helps them revise multiple times as the content can be re-watched or re-read multiple times unlike in a classroom environment.

Another example could be when an assignment or project must be submitted. The student has many options open to him on the internet with an ocean of similar projects that can easily help set precedence to his project. There are websites which provide free expert reviews of student's projects and help them create it from scratch, from building blueprints onward and even for free. This was obviously much more economic compared to when we they had to purchase a book for the same purpose.

Yet another instance could be the paramount help e-learning is during competitive exams and university tests. Instead of spending many hours at tedious and monotonous lectures, there are interesting and fun ways of learning concepts via puzzles, questionnaires, online group activities etc. on the internet which have been proven to be more effective due to shortcut methods available, even if it is a last-minute preparation. They also have access to large databases of question papers to refer and study from. And based on each student's caliber, there are specialized test questions available. The mock tests online also help give them an approximate idea of their preparation level and an expected score range.

So, we can see that a student's learning curve is vague but still dependent on the method of e-learning that he chooses. To make a wise choice he must pick the method that is most comfortable to him and suited to his understanding and grasping power. The different choices he has include:

**2.1. STATIC LEARNING**

This is a one-way street where the student's only interaction with the web is to access the disseminated and stored information or content. This includes but is not limited to PowerPoint slides, E-books, question papers, videos, recorded lectures and related material. On picking up the relevant content from sources such as WhatsApp class groups, University-Notes portals etc., the student conducts domestic self-study with the help of these collectives.

**2.2. DYNAMIC LEARNING**

This usually consists of a regular online community that is balanced by two poles: The 'teacher' and the 'pupil'.

This is further divided into:

- Private tutoring: The teacher imparts knowledge and ideas to be learned and explains them to the pupil in a one-on-one session.

- Group E-Learn Meets: Two or more people who are well versed in a particular common area of interest get together and exchange ideas to help each other out. Both are equal participants in this exchange.

- The Online Classroom: A qualified individual schedules a class and delivers a lecture to an entire class of interested individuals live via Skype, Blackboard, FaceTime, Google Plus Hangouts, Udacity etc., and also conducts a questionnaire session for doubt discussions.

Most of the higher education institutes and universities in India are adopting the online classroom method. Students are getting on to the E-learning journey either by choice to skill up in new areas or as part of their university curriculum. Once he selects what is personally best for him, he is all set to go on his E-learning journey.

## 3. ALONG THE WAY

The student starts walking down his selected road, but he encounters a lot of deviations and obstacles.

With universities enforcing online tests and online submissions as a focal part of education, there is the rising threat to authenticity. While e-learning enabled students to work time-effectively, it still had its shortcomings with respect to college level testing. Students could always duplicate content from online easily using plugins for every type of testing software. This is majorly due to the general technical drawback of the online world: there was always a way to break in, hack and alter the functioning of most systems easily. This has made it increasingly easy for students to indulge in malpractice since every step to performing such unethical hacks are present openly online paradoxically.

So, in an era where cybercrimes are a matter of fact, online testing might not be a completely apt representation of a student's true potential and proficiency.

But since it's our best bet for development and still more effective than traditional learning, we move on to analyzing it further. At university level, to be successful at web learning, there are multiple factors that contribute apart from the students learning capacity and tutor/instructor:

- The student's attitude.
- The content organization or knowledge base.
- An interactive tool or the web-based platform used as media for communication, which is more important than the content itself.
- Effective feedback or assessment module from students who are currently taking the web-based course or have done so in the past.

The modern E-learning system at university levels is described in Figure1. With help of industry experts and content creators, staff will be able to provide course content. The interactive learning console helps navigating to the topic that student wants to learn. The Intelligent module feeds the content based on the students understanding and feedback. The admin and staff also feed in the assessment questions. Randomized set of questions provided to students for evaluation of their learning. Staff get report on student progress and feedback for further review.

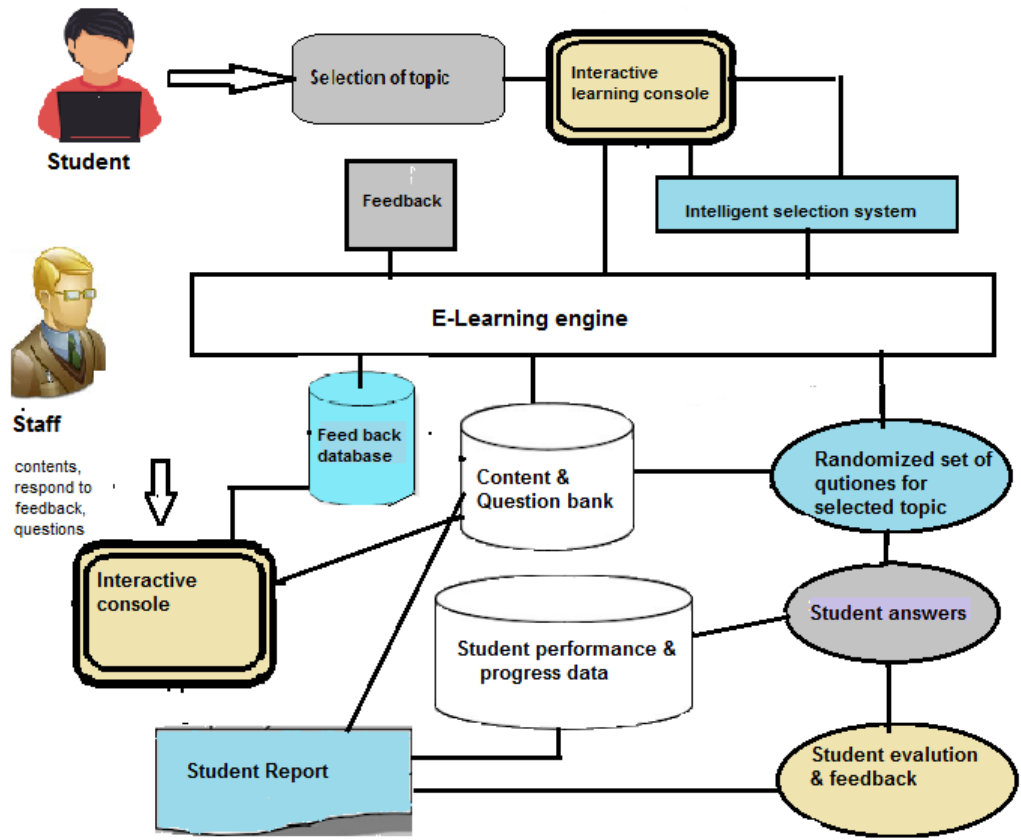

Figure 1: Process involved in e-learning at university level

Many universities are launching distance education programs using web-based learning and online libraries. With India being the current third largest online market for education in the world, it has launched NPTEL-a Govt. initiative and website where regular courses from reputed Indian institutions hosted for free. One such similar but private e-learning instance in India is eLab.

## 3.1. THE REVIEW OF eLAB

SRM university's eLab is an auto-evaluation tool for learning programming. eLab helps learners to practice and acquire programming skills. Sophomore students are warmed up to the new coding languages they would be learning through a semester using this tool.

Once the student logs in, he gets an effective pictorial representation of his question completion status:

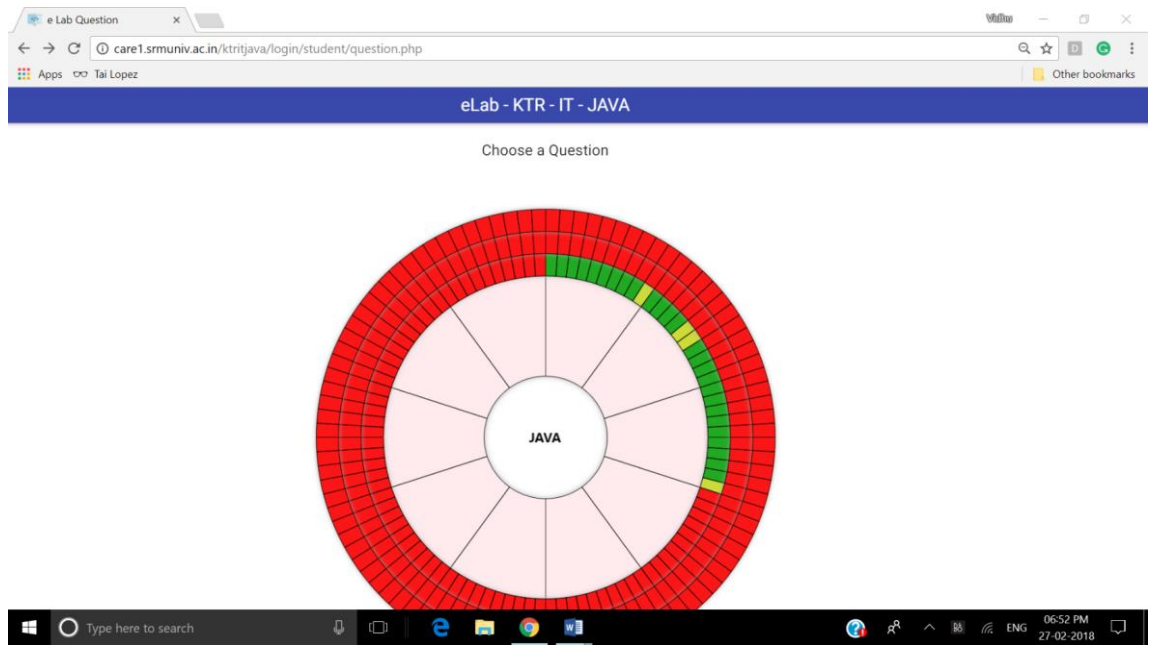

Figure 2. The eLab completion status wheel.

The green slots represent the questions executed successfully and the yellow and red would represent the attempted but incomplete and un-attempted questions respectively.

With a step-by-step approach, it levels up in competitiveness with every question. There are several sections present in the tool, to cater to all the subjects the student will be completing in the current semester. Each section within a subject would pertain to a chapter of the subject. Thus, all areas of the student's study, would be thoroughly tested. Each student is required to complete a set of programs whose problem statements will be defined to them along with example test cases (as shown in the screenshot below). The student is required to analyze the question and code to satisfy the test cases. This system executes the written code and returns a result or highlights the issues with the code if insufficient. The system is also designed to award a completion score in percentage. If the code was ideal and satisfied all conditions, it recorded a 100%. If the code worked in favor of the question but would not satisfy all conditions, it recorded a 33,50 or 75 percent based on how much improvement the code needed. Finally, an average of all these scores makes the final cut to be added to the student's grade point average(GPA).

Initially the access to this tool was limited to devices connected via the university's private network. Now it has been made available for use from any network. So now students can use it from anywhere including home, to complete their graded college work and submit it as well. But the list of pros doesn't stop there. Teachers and respective faculty gain well from this tool too. They can check the student's progress at any given time, review & reframe the question set, and download a progress report for all students that registered under them.

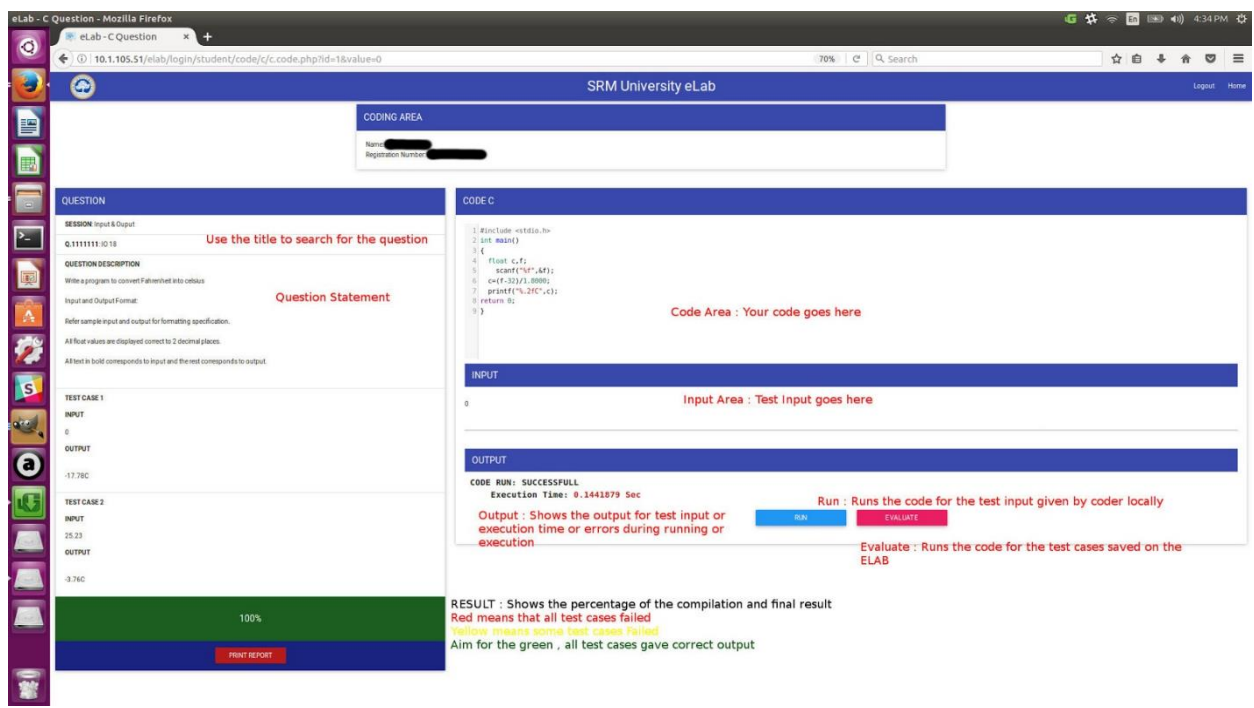

Figure 3. The eLab user interface

The tool's access specifications are sealed in the backend, due to which the students cannot change of edit their grades or scores. Additionally, since the same software is used for multiple subjects, it works well to satisfy the university's needs while constantly acting as an effective learning tool for the student learners. Though the tool has a very good infrastructure for web learning, the content must match real-world problems, else it would defeat the purpose of learning programming. With students dealing with university enforced tools with a lot of loopholes, they lag behind when they are sent into the industry.

## 4. ZOOMING IN ON THE GAP

With the advent of e-learning in colleges, the universities curriculum started to get greased and easy. But this was only one side of the learning curve because on the other end of the curve, were graduates passing out of the universities with no knowledge of the industry whatsoever. They were incomplete products whom the companies seem to devour and shape to suit their specific needs.

According to HR reports, around 1.5 million engineers are released into the job/working market every year. But only about 25% of them are directly employable from the college campus. So, this goes to say that employers today look for more than just the basic skill set in their employees, they look for a long-term relationship with the organization. Which is why they focus on reskilling. Graduate students are required to unlearn and learn to fit into the IT world. So, the educators need to start thinking ahead of the curve.

This is where companies like Embibe and Simplilearn enter the picture with their courses for mid-level professionals. They help create a smooth segway from college to industry. Many businesses empires have been built around helping this university-to-company transition: Coursera, Udacity, edX etc. These e-learning websites primarily focus on delivering certified industry level courses which are now even being prescribed by companies itself! MNCs like Google, Microsoft and IBM now ask their employees to refer to such common e-learning portals instead of conducting intense and tedious training programs. A Brandon-Hall Study proved that learning online required almost 40% to 60% less employee time than learning the same topic in a traditional classroom setup. For instance, when IBM implemented an e-learning program in their company, its employees learned five times more without increasing their time spent in training.

When the freshers felt unprepared for the market, they ran to these web-based systems for a revival. On completion of these online courses from Coursera etc., they would've added jewels of value to their CVs which immediately helped them get hired and have also eased their transition into the walls of the industry since companies preferred such candidates over newer unprepared college minds.

## 4.1. HOW THE GAP COULD BE HANDLED

- Course content for universities should start to be: prescribed by or delivered directly from industry partners or companies, as the constant updates that are done by college seem insufficient. Therefore, the responsibility comes upon the system to update the syllabus and the pedagogy based on the society's demands.

- Universities should use professional content developers who have the industry experience.

- Web-based learning systems should consider using Artificial Intelligence. The tools should be smart and more interactive in nature. It should learn by itself and adjust the complexity of the delivered course based on the interaction and feedback of the students.

- The evaluation tools should predict the student understanding level and adjust the complexity level and provide the set of questions dynamically based on their response and award accordingly.

- Colleges should create space for students with a more direct inclination to the industry and inculcate that as part of their graded courseware, as outcome-driven academic systems flourish better.

- More web-based courses should be introduced into university curriculum to help industry freshers since they come with no experience of the workplace pressure, workload volumes or co-ordination required at companies at a global level. They need to be thought how to unlearn what they have collected in domestic college environments and learn what real-time systems require.

- Industries can create video content about the process and technology that they expect from the students to be employer ready and publish education material to students and universities.

- Industries should update the happenings of various conferences to students. This would enable the students to align their reach projects, self-learning etc. using the web-based education offered.

- While offering eLearning feedback, it's important to only concentrate on behaviors, actions, or skills that can be changed or improved upon.

## 5. IN CONCLUSION

With the rigorously increasing E-learning appetite, the education system has improved by leaps and bounds as compared to what it was in the past decade. But ever since the comfort of studying on your own couch was introduced, the lazy students sense the main turbulence of studying at their own pace - the procrastination. Students need to have a highly motivated self-learning drive since web-based learning eliminates travel, timings and sometimes deadlines too and makes it all too flexible. But we cannot deny the fact that both the economy and literacy levels have gone up since the advent of web-based e-learning.

Furthermore, tailored courses with flexible completion timing do enhance majority of students' inclination toward learning. The stepping stool into the IT industry has become much larger with e-learning. More and more companies are starting to recognize this trend. It is due to the abstraction feature of web-learning that, from a sea of unrelated information we get straightforward and tailored content delivered to us for learning. Thus, employees are now able to deliver better and effective results at within shorter time spans. So, web-learning has contributed to nothing but good to the technical revolution. But with just a few more changes and guideline enforcements, in the foreseeable future, e-learning would have revamped the book era entirely and become the new worldwide education standard. Universities would be technology enabled with a much leaner infrastructure. And in this changed paradigm, the learner is the new center of the knowledge-universe with his personalized learning path.

**AUTHOR**

Vidhu Mitha Goutham is a second-year student pursuing Engineering in Information Technology at SRM University in Chennai, India with a keen interest in voicing to the online community.